\documentclass[review]{elsarticle}

\usepackage{mathtools}

\journal{Adv. Water Ressources}

\begin{document}

\begin{frontmatter}

\title{Pore-scale simulations of drainage in granular materials: finite size effects and the representative elementary volume}

\author{Chao Yuan, Bruno Chareyre, F\'elix Darve}
\address{Univ. Grenoble-Alpes, 3SR, F-38000 Grenoble, France and\\ CNRS, 3SR, F-38000 Grenoble, France}




\begin{abstract}
A pore-scale model is introduced for two-phase flow in dense packings of polydisperse spheres. The model is developed as a component of a more general hydromechanical coupling framework based on the discrete element method, which will be elaborated in future papers and will apply to various processes of interest in soil science, in geomechanics and in oil and gas production. Here the emphasis is on the generation of a network of pores mapping the void space between spherical grains, and the definition of local criteria governing the primary drainage process.
The pore space is decomposed by Regular Triangulation, from which a set of pores connected by throats are identified.
A local entry capillary pressure is evaluated for each throat, based on the balance of capillary pressure and surface tension at equilibrium. 
The model reflects the possible entrapment of disconnected patches of the receding wetting phase. 
It is validated by a comparison with drainage experiments. In the last part of the paper, a series of simulations are reported to illustrate size and boundary effects, key questions when studying small samples made of spherical particles be it in simulations or experiments.
Repeated tests on samples of different sizes give evolution of water content which are not only scattered but also strongly biased for small sample sizes. More than 20,000 spheres are needed to reduce the bias on saturation below 0.02. Additional statistics are generated by subsampling a large sample of 64,000 spheres. They suggest that the minimal sampling volume for evaluating saturation is one hundred times greater that the sampling volume needed for measuring porosity with the same accuracy.
This requirement in terms of sample size induces a need for efficient computer codes. The method described herein has a low algorithmic complexity in order to satisfy this requirement. It will be well suited to further developments toward coupled flow-deformation problems in which evolution of the microstructure require frequent updates of the pore network. 
\end{abstract}


\end{frontmatter}


\section{Introduction}
\label{sec:introduction}
Understanding two-phase flow in granular media and the coupling with deformations of the granular skeleton is of great importance in many areas of engineering and science. This includes transfers in soils and associated phenomena such as desiccation cracks, swelling and slopes instabilities, various oil recovery techniques or the extraction of methane hydrates from sea bed sediments. Microscale imaging techniques together with pore scale numerical models are promising tools for gaining insight into the governing mechanisms of two-phase flow in such systems. However, both experimental techniques and computational methods have severe limitations in terms of sample size, which raises questions about possible finite size effects. On the modelling side, this difficulty is amplified when coupled flow-deformation processes are simulated, since time integration implies repeated executions of the flow solver in an ever changing pore geometry, hence even smaller problem sizes to keep computati
 onal costs acceptable. The aim of this work is twofold: to introduce a pore-scale method enabling high speed simulation of drainage in dense sphere packings and to evaluate finite size effects in such systems. The case of a deforming skeleton is not explicitly tackled yet, but this work is clearly meant for a step in this direction. For this reason the method has to be compatible with the direct simulation of deforming granular structures.


There are several approaches at different scales available for simulating two-phase fluid flows.
Macro-continuum scale models are used in most field-scale applications. They are based on empirical relations describing, namely, capillary pressure - saturation ($P_c-S_w$) curves and their evolution with strain, relative permeability, and effective stress. Such methods have acceptable computational costs for large problems, but the empirical laws therein have well known issues. Namely, hysteretic effects are very difficult to model and an accepted effective stress framework is still missing.

The micro-continuum scale methods, which include lattice Boltzmann (LB) method, volume of fluid (VOF) method, smoothed particle hydrodynamics (SPH) method and level set method, do not rely on such empirical relations. They are promising approaches for getting accurate results at very small scales and gaining understanding of the phenomena observed at the macroscale. However, they have high computational cost.

Pore-network models introduce an intermediate scale at which pore bodies are identified. They enable the simulation of larger domains compared to micro-continuum models, with much fewer assumptions than the macro-continuum models. They idealise the porous medium as a network of pore bodies connected by narrow throats. Pore-network modelling was pioneered by Fatt (see \cite{Fatt1956a} and companion papers), who derived $P_c-S_w$ curves of primary drainage and computed pore size distributions in a network of interconnected pores. Since then, a number of different researchers have contributed to the current understanding of two-phase flow using pore-scale models. The model we present herein differs from previous pore-network models in some aspects, but the general methodology is very similar.

Many networks are based on regular lattices. Typically, squared lattices with a coordination number of four in two-dimension (2D) or cubic lattices with a coordinate number of six in three-dimension (3D)\cite{Chandler1982, Berkowitz1993, Stark1991}.
The shape of the pores have been approached by regular geometries ({\it e.g.}, cubic \cite{Joekar-Niasar2010b,Raoof2013} or spherical \cite{Koplik1985}) and the shape of the pore throats by cylinders with various cross-sectional shapes ({\it e.g.}, circular \cite{Dias1986, Koplik1985} or triangular \cite{Al-Gharbi2005}) or with parallel pipes\cite{Hughes2000,Joekar-Niasar2010b}.
Angular cross sections have been proposed by some authors to reflect the phenomena of corner flow and the crevices occupied by the wetting phase.
Statistically representative pore networks of this kind can be generated to represent real porous samples \cite{Khaksar2013,Rostami2013,Nikooee2014}.

Other work focuses on mapping directly the pore space of real granular materials, a problem pioneered by Bryant and Blunt \cite{Bryant1992} (see also \cite{Bryant1993a, Bryant1993b}), who constructed a network mapping an experimental specimen of packed mono-sized spheres.
As imaging techniques reach smaller and smaller scales \cite{Culligan2004, Khaddour2013}, there is a growing interest in this problem and many pore-network extraction algorithms are being developed. They include the multi-orientation scanning method\cite{Zhao1994}, medial axis-based algorithms\cite{Lindquist1996, Sheppard2005, Prodanovic2006b}, Delaunay/Voronoi diagram-based methods\cite{Bryant1992, Oren2003}, or the method of maximal ball\cite{Silin2003, Silin2006}). 

Solving coupled flow-deformation problems similarly requires to map a network directly on a given set of solid particles. Moreover mechanical coupling requires a direct and explicit link between the network geometry and the positions of the solid grains, and the computational cost of updating the network should be kept as small as possible. It makes the Delaunay/Voronoi methods best candidates since they introduce a simple duality between the solid objects and the void space. The very few existing models coupling a pore-network with a deforming material also adopted this methodology (in two-dimensions \cite{Jain2009,Kharaghani2012}, hence mainly qualitative).
A three-dimensional pore-scale approach termed \textit{PFV} \cite{Chareyre2011} has been proposed to effectively solve flow problems with a single fluid phase. Therein, the fluid flow was modelled using a pore-scale finite volume scheme (PFV) which shares many features with conventional pore-network methods. The method has proven effective in the context of coupled flow-deformation problems \cite{tong2012,catalano2014,scholtes2015}. 
In this paper we propose a new model extending the flow model of \cite{Chareyre2011} to quasi-static two-phase flow, as a first step toward coupling two-phase flow and deformation. Beside accuracy, merits of the model are its computational efficiency and its ability to deal with poly-disperse spheres. The coupling with a deformable solid skeleton will be addressed in future publications.

All the measurements or simulation may be sensitive to the problem size and boundary conditions, but this is particularly true when the said sample is made very small because it has to fit in a tomography apparatus \cite{Baveye2002} or because the computational resources are limited \cite{Hilpert2001}. This is a source of difficulty for the validation of pore scale methods when some details of an experimental setups cannot be reproduced exactly or are simply not known. Anticipating further works in this direction, the first application of our model - and second part of this paper - is a systematic analysis of size effects and boundary effects in small-sized packings of spheres.
%
%
The structure of the paper is as follows.
In Secs.\ref{sec:pore-scale_network} and \ref{sec:drainage_model}, we first employ a decomposition technique to build the pore-network and we introduce the governing equations of drainage.
To accommodate different assumptions and experiment situations, various phase trapping options and boundary conditions are implemented.
In Sec.\ref{sec:comparison_with_experiment}, the model is validated by a comparison with quasi-static drainage experiments.
In Secs.\ref{sec:REV} and \ref{sec:side_boundary_size_effect}, repeated simulations of drainage in random packings are reported to study finite size effects. Namely, we discuss the effects of sample size, the statistics of saturation obtained by subsampling, and the role of boundary conditions and aspect ratio on capillary pressure-saturation relationships and phases distribution. 

\section{Pore-scale network}
\label{sec:pore-scale_network}
We consider materials in which the solid phase can be seen as a random dense packing of poly-disperse spheres. Such packings will be generated with the DEM method\cite{Yade-DOC}.
The network representation of the pore space is obtained in three dimensions by using the Regular Triangulation method, in which the tetrahedra define pore bodies and the facets correspond to the pore throats. The algorithm follows \cite{Chareyre2011} and it is only briefly summarised hereafter. 

\begin{figure}
 \centering
\includegraphics[width=0.8\columnwidth]{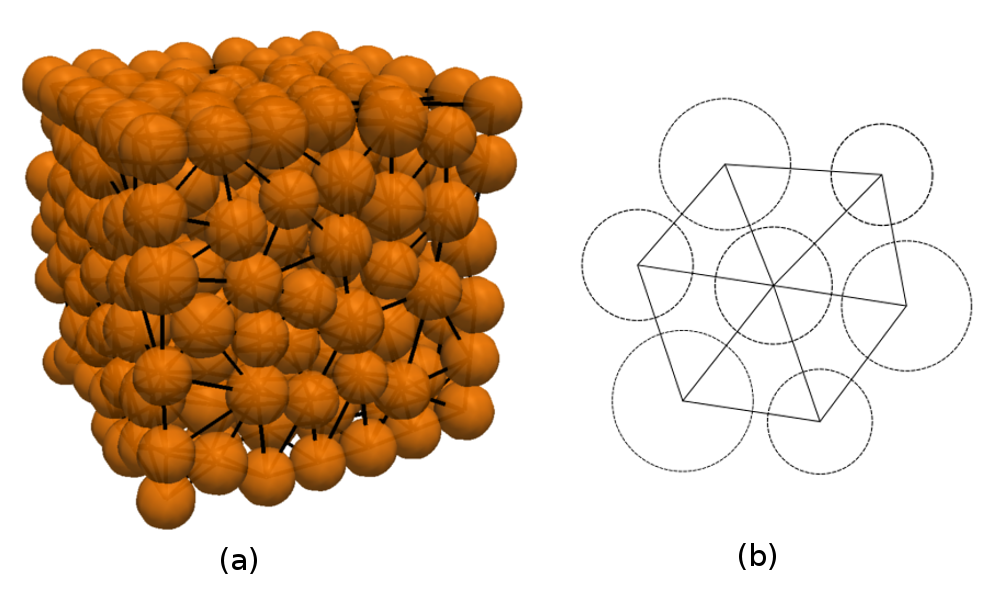}
\caption{Regular triangulation in three dimensions (a) and two dimensions (b).}
\label{fig:triangulation}
\end{figure}

\begin{figure}
 \centering
\includegraphics[width=0.8\columnwidth]{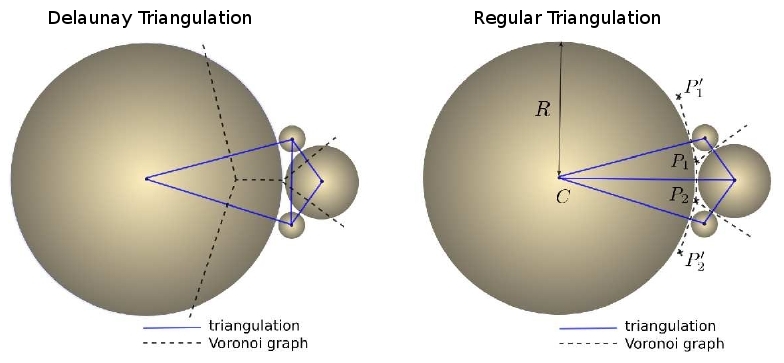}
\caption{Comparison of triangulations and their dual Voronoi graph in two dimensions. (Left) the dual graph of Delaunay triangulation has branches inside discs; (right) the dual of the Regular Triangulation has all branches in the pore space.}
\label{fig:delaunay}
\end{figure}

Regular Triangulation (also known as weighted Delaunay triangulation or power diagram) generalises the classical Delaunay triangulation to weighted points, where the weight accounts for the size of each sphere\cite{Edelsbrunner1996}. Typical examples are shown in Fig.\ref{fig:triangulation}.
The dual Voronoi graph of regular triangulation (also known as Laguerre graph or radical Voronoi graph) is based on radical planes and it is entirely contained in the void space. 
This is an appropriate feature to describe the flow path within the pore space, as opposed to the classical Delaunay/Voronoi graphs (see Fig.\ref{fig:delaunay}).

\begin{figure}
 \centering
\includegraphics[width=0.8\columnwidth]{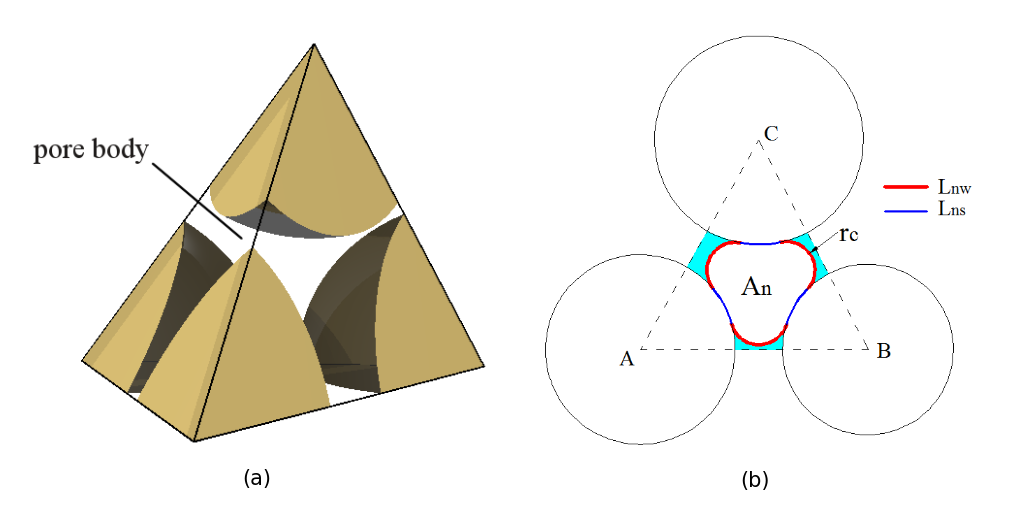}
\caption{Pore geometry. (a) A pore defined by a tetrahedral element. (b) Geometry of a pore throat.}
\label{fig:poreGeometry}
\end{figure}

Based on this decomposition, a pore is surrounded by four solid spheres whose centres are the vertices of the corresponding tetrahedron. 
The volume of the pore body corresponds to the irregular cavity within the tetrahedron (see Fig.\ref{fig:poreGeometry}(a)).
The shape of a pore throat is defined by the cross sectional area extending within a tetrahedral facet (Fig.\ref{fig:poreGeometry}(b)). The throat does not enclose any volume, but it will play a key role when defining the entry capillary pressure of an invading non-wetting phase (NW-phase).

Since each pore corresponds to a tetrahedron, it has four neighbours, resulting in a lattice of connectivity equal to four.
Relatively similar networks can be found in other models \cite{Mason1971,Mason1995,Bryant1993a,Gladkikh2003}, yet the decomposition techniques therein are restricted to uniform particle sizes by the choice of Delaunay triangulation. Regular triangulation extends the approach to poly-disperse spheres. Its mathematical definition is limited to geometrical arrangements of non-overlapping or moderately overlapping spheres. More precisely the maximum overlap is when the centre of one sphere enters another sphere, in this occurrence the regular triangulation would be undefined. Since repulsive forces at contacts prevent such overlaps when the assembled spheres represent solid grains, the regular triangulation that is adopted is always defined.

\section{Drainage model}
\label{sec:drainage_model}
\subsection{Local rules}
\label{subsec:local_rules}
In the absence of gravity the movement of immiscible phases occurs in different regimes distinguished by the relative contribution of viscous stresses and surface tension. The balance between the two depends on two dimensionless numbers, the viscosity ratio $M$ and the capillary number $C_a$, 
\begin{equation}
 \label{equ:viscosity_ratio_capillary_number}
M=\frac{\mu _{inv}}{\mu},  C_a=\frac{\mu v}{\gamma}
\end{equation}
where $\mu_{inv}$ is the viscosity of the invading phase, $\mu$ is the viscosity of receding phase, $v$ is the average velocity of the receding phase, and $\gamma$ is the interfacial tension between the two fluid phases(\cite{Lenormand1988,Lenormand1990}).  
The limit of ``quasi-static'' flow corresponds to $C_a=0$, a situation in which the viscous effects can be neglected.

The model we propose aims at simulating the slow primary drainage of air-water systems, or more generally non-wetting/wetting (NW-W) systems.
We assume a quasi-static regime and a perfect wetting of the solid (S-phase) by the wetting phase (W-phase). Consistently, the fluid pressure is piecewise uniform in every set of pores occupied by a certain phase and connected through this particular phase, i.e. one fluid cluster has only one fluid pressure (it applies for both the W and the NW phases).
The local pressure can differ from the reservoir pressure of the same phase only in a group of pores disconnected from the reservoir.

The drainage process is controlled by the capillary pressure $P^c$, {\it i.e.}, the pressure difference between the NW-phase and W-phase: $P^c=P^n-P^w$. In quasistatic flow, the invasion of a pore body can be seen as an instantaneous event in which the interface moves from one throat to the next ones by a so-called \textit{Haines jump}. Thus the interfaces are always located near the throats in the simulation, practically. 
When a pore saturated by the W-phase is adjacent to another pore already invaded by the NW-phase, the stability of the W-NW interface at the corresponding throat depends on the entry capillary pressure $P^c_e$ associated to the throat (see next section). If $P^c>P^c_e$ then the pore is invaded. 
In principle, a certain amount of the receding W-phase can be left behind in the invaded domain in the form of disconnected pendular rings \cite{Scholtes2009a,Scholtes2009b}. 
At this stage we neglect the volume of such rings when determining the total volume of each phase. Neither do we consider the presence of W-phase in the corners of the throats (a situation sometimes considered for prismatic pore throats) as it would make little sense in sphere packings.
To sum up, the pore space is entirely contained in the pore bodies and the saturation of one pore is simply binary, {\it i.e.}, it equals 0 or 1 depending on which phase is present.
Obviously, some real situations may differ significantly from such idealisation. The residual saturation in particular may be modified by imperfect wettability, leaving some pores only partially drained, or by viscous effects - especially for high viscosity fluids such as oils. The comparison with experiment in section 4.2 suggests that the simplifications may be acceptable for a first approach of some air-water systems. Further model refinement will be necessary for simulating more general conditions.

\subsection{Determination of entry capillary pressure}
\label{subsec:determination_of_Pc}
In fluid statics, a relationship between capillary pressure, $P^c$, interfacial tension, $\gamma^{nw}$, and mean curvature of the NW-W interface, $C$, is given by the Young-Laplace equation
\begin{equation}
 \label{equ:Young_Laplace_equation}
 P^c = 2C\,\gamma^{nw} .
\end{equation}
$C$ can be expressed in terms of the principal radii of curvature of the meniscus ($r_1$ and $r_2$) by
\begin{equation}
 \label{equ:curvature_equation}
 2C=(\frac{1}{r_1} +\frac{1}{r_2})
\end{equation}
This is a starting point for defining $P^c_e$, yet $r_1$ and $r_2 $ are difficult to define precisely for an interface near a pore throat of complex geometry. Approximations are necessary. We propose to determine $P^c_e$ based on MS-P (Mayer-Stowe-Princen) method, which employs the balance of forces on the NW-W interface (\cite{Mayer1965,Princen1969b}). The balance reads
\begin{equation}
 \label{equ:force_balance}
 \sum F(P^c)= F^p(P^c) + T^{\gamma}(P^c) =0
\end{equation}
where $F^p$ is the capillary pressure acting on the interface and $T^{\gamma}$ is the total tension force on multi-phase lines. $P^c_e$ is the value of $P^c$ such that $\sum F(P^c)=0$.
This method for determining $P^c_e$ follows \cite{Ma1996, Prodanovic2006a, Joekar-Niasar2010a}. Therein, the MS-P method is applied to cylindrical throats.
Our situation is more complex since the cross sectional shape is changing along the flow path. By employing the MS-P method we {\it de facto} assume that $P^c_e$ is the same as in a cylindrical throat tangent to the solid phase at the narrowest cross section, an assumption which will be evaluated in section \ref{sec:comparison_with_experiment}.
For completeness, we recall the generic aspect of the MS-P method hereafter. 

Figure\ref{fig:poreGeometry}(b) shows the typical geometry of a pore throat and the parts occupied by the different phases and interfaces.
If $P^c$ increases the region occupied by the NW-phase grows, pushing the W-phase further toward the corners of the throat.
The longitudinal curvature of the W-NW interfaces is supposed to approach zero as $P^c$ approaches $P^c_e$\cite{Joekar-Niasar2010a}, {\it i.e.}, $r_1\to+\infty$ in eq.~\ref{equ:curvature_equation}.
Assuming that both phases pressure are uniform around the throat, the remaining cross-sectional curvature $r_2$ must take the same value for all three W-NW interfaces (based on Eqs. \ref{equ:Young_Laplace_equation}-\ref{equ:curvature_equation}). This value is denoted by $r_c$ and it is related to the entry capillary pressure
\begin{equation}
 \label{equ:capillary_pressure}
 P^c_e = \frac{\gamma^{nw}}{r_c}
\end{equation}

For a geometry of pore throat as defined in Fig.\ref{fig:poreGeometry}(b), the forces acting on the interface are the force coming from the capillary pressure exerted on that part of the cross section occupied by the NW-phase
\begin{equation}
 \label{equ:entry_force}
 F^p = P^c A_n,
\end{equation}
and a force coming from surface tension on the multiphase lines
\begin{equation}
 \label{equ:tension_force}
 T^{\gamma} = L_{nw} \gamma ^{nw} + L_{ns} \gamma ^{ns} - L_{ns} \gamma ^{ws},
\end{equation}
where, $A_n$ is the NW-phase area inside the throat's section, and $L_{nw}$ and $L_{ns}$ are the total lengths of NW-W interfaces and NW-S interfaces respectively. 
The multiphase interfacial tensions, $\gamma ^{ns}$, $\gamma^{ws}$ and $\gamma ^{nw}$ have a relationship with contact angle $\theta$, defined by Young's equation,
\begin{equation}
 \label{Young's equation}
 \gamma ^{ns}- \gamma^{ws}= \gamma^{nw} \cos \theta
\end{equation}
Then Eq.\ref{equ:tension_force} gives
\begin{equation}
 \label{equ:tension_force_simplify1}
 T^{\gamma} = (L_{nw}+L_{ns}\cos\theta)\gamma^{nw} 
\end{equation}

All terms of Eq.\ref{equ:entry_force} and Eq.\ref{equ:tension_force_simplify1} can be expressed as functions of $r_c$ (see Appendix), so that the equilibrium equation Eq.\ref{equ:force_balance} is an implicit definition of $r_{c,e}$, the value of $r_c$ for which the equation is satisfied (noting that $\sum F$ is a monotonic function of $r_c$):
\begin{equation}
 \label{equ:force_balance_rc}
 \sum F(r_c)= F^p(r_c) + T^{\gamma}(r_c)=0
\end{equation}
Solving the equation numerically gives $r_{c,e}$. In turns $P^c_e$ can be determined using Eq.\ref{equ:capillary_pressure}.

\subsection{Drainage and entrapment of W-phase}
\label{subsec:drainage_and_trapping}
\begin{figure}
 \centering
\includegraphics[width=0.4\columnwidth]{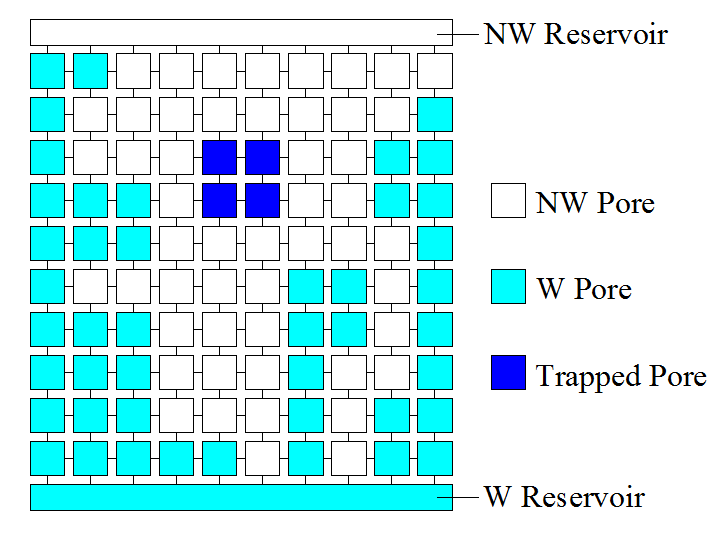}
\caption{Illustration of boundary conditions, NW-phase invasion and W-phase trapping in the network (in 2D mapping for clarify).}
\label{fig:network2D}
\end{figure}

In order to explain the invasion logic of our model we represent the 3-D network using a 2-D lattice mapping (see Fig.\ref{fig:network2D}).
Pore bodies and throats are represented by squares and linear connections respectively.
Different flags are assigned to the pores to reflect the individual state of saturation (0 or 1) and whether a particular pore or a group of pores is directly connected to one of the main reservoirs. A search algorithm is employed for updating those states during invasion.

Initially, the sample is saturated, and the top and bottom boundaries are connected to NW and W reservoirs, respectively. 
The effect of gravity is ignored.
Drainage starts by increasing the NW-phase pressure $P^n$ of the NW reservoir while the W-phase pressure $P^w$ in the W reservoir is kept constant (thus increasing $P^c$).
A search is executed on the pore throats which separate the phases. The throat with lowest $P^c_e$ is where the first displacement of the interface will occur (Haines jump), leading to the invasion of a first pore by the NW phase.
As soon as this pore is invaded the NW phase reaches new throats, possibly triggering a recursive cascade of Haines jumps and invading more than one pore for one single value of applied $P^n$, until no more throats satisfy $P^c_e<P^c$. It leads to discontinuous changes of the W-phase content which have been also observed in experiments{\cite{Culligan2004}}.
When the simulation reaches the new stable configuration for a certain applied $P^c$, the state flags are updated for the next step of drainage.

As the NW-phase is invading, the W-phase may form clusters of pores which are disconnected from the W reservoir. In order to identify these entrapment events, a dynamic search algorithm is employed after each drainage event.
We assume that there is no film flow or evaporation in the model, thus the disconnected regions remain saturated by a fixed amount of the W-phase throughout subsequent increases of $P^c$.
Though acceptable for sufficiently fast drainage, this simplification may lead to slightly overestimate the W-phase content of samples subjected to $P_c$ for longer periods of time.

The W-phase is assumed to be incompressible, so that the geometry of the W-phase and NW-W interfaces for disconnected regions remains unchanged throughout the next steps of drainage.
According to Eq.(\ref{equ:Young_Laplace_equation}), $P^c$ will also remain the same because of unchanged NW-W interfacial curvature. 
Therefore, $P^w$ in trapped W-phase must grow along with $P^n$. Since the disconnections of different regions may happen at different times of the drainage, every disconnected region has its own local value of $P^w$ ultimately (consistently with \cite{Harris1964} for instance).

\subsection{Boundary conditions}
\label{subsec:boundary_conditions}
We define boundary conditions that mimic realistic drainage tests on finite-sized samples.
The particles are packed in between rigid walls.
From a mathematical point of view the rigid walls are represented as spheres with near-infinity radii, handled as ordinary spheres by the triangulation algorithm.

We define the top and bottom layers of pore units as connected to the NW and W reservoirs, respectively.
They will remain in the initial state throughout the drainage simulations ({\it i.e.}, constantly occupied by the same phase).
Correspondingly, the calculation of saturation will not involve the volume of these boundary pores.

Special attention has been payed to the connectivity of side boundary pores and throats (located between a vertical wall and the first layer of spheres along this wall). It will be seen that they can play a dominant role in the drainage process, in agreement with \cite{Chandler1982}.
In the model, we can decide whether such pores should be available for invasion (``open'') or just disregarded and not participating to the system (``closed''). 
The latter case is hardly related to any realistic test condition although it may mimic the effect of a rough boundary. It is introduced to enable interesting comparisons, mainly.
The calculations of saturation are adapted to the different scenarios and exclude the boundary pores when they are closed.

\subsection{Implementation}
The network generation has been implemented in C++ \cite{Chareyre2011} as part of the open-source code Yade-DEM \cite{Yade-DOC}, with the help the geometric algorithms library CGAL \cite{Boissonnat2002}. CGAL provides very efficient algorithms for regular triangulation. 
The network generation (initially devoted to one-phase flow) has been complemented with the set of functions for the determination of entry capillary pressure and for updating the state flags and connectivity flags after each event. As a whole, this pore-scale model is freely available as part of the open-source discrete element code Yade-DEM and the results presented in the next sections can be reproduced independently.

\section{Comparison with experiments}
\label{sec:comparison_with_experiment}
\subsection{Numerical setup}
\label{subsec:numerical_setup}
In this section, we verify the pore-scale model by comparing the simulation results with experimental data from a quasi-static drainage experiment in a synthetic porous medium{\cite{Culligan2004}}.
The measurements were done on packed glass beads, contained in a column of 70 mm in length and 7 mm in diameter.
The particle size distribution (PSD) of the glass beads was as in Table.(\ref{tab:PSD}), the porosity is 0.34.
The drainage was carried out by pumping water out of the porous medium.
X-ray tomography was used to image only a small part of the column (a cubic box of size 5 mm approximately) to determine capillary pressure-saturation ($P^c-S^w$) relationships.
\begin{table}
\begin{center}
\begin{tabular}{*{5}{c}}
\hline\noalign{\smallskip}
Weight(\%) & Diameter(mm) \\
\hline\noalign{\smallskip}
30 & 1-1.4\\
35 & 0.850\\
35 & 0.600\\
\hline
\end{tabular}
\end{center}
\caption{\label{tab:PSD} Particle size distribution of the experiment}
\end{table}

Due to some properties of the regular triangulation, generating the pore-space decomposition for a column of circular cross-section (as in the experiment) would have been excessively complex in the present state of the algorithms. Instead, cuboid samples are generated, with the same average properties as the experiments in terms of PSD and porosity. The simulated packings are connected to the NW reservoir at the top, and to the W reservoir at the bottom, as in figure \ref{fig:network2D}. The drainage process is simulated by imposing a progressive increase of the NW-phase pressure in the NW-reservoir (and keeping the W-phase pressure constant). There is no gravity in the model. Since gravitational problems are equivalent to non-gravitational ones if piezometric pressure is used in lieu of absolute pressure, it does not induce a loss of generality as long as gravity does not modify the curvature of the interfaces (a rather good approximation for air-water systems with grain size below 
 1mm \cite{Pitois2001}).

Due to capillary fingering, the situation at the boundaries of the window accessible by tomography is difficult to define precisely.
In the simulation, it may be assumed that only a few large pores of the boundaries are connected to the invading phase reservoir (reflecting fingering in that part of the column not scanned, with some fingers reaching the scan region). Conversely, it may be assumed that all pores of the boundary associated to the NW-phase reservoir are occupied by the NW-phase. The two variants could lead to significant differences in the results in some circumstances \cite{Joekar-Niasar2010a}. In our case, preferential invasion along the boundaries (an effect which we will discuss in details in section \ref{sec:side_boundary_size_effect}) reduces significantly the influence of the reservoir connectivity. For the sake of simplicity we use the last assumption, {\it i.e.}, uniform boundary conditions on the top and bottom faces of the box.

For one simulation, a cubic box of size 5.0 mm$\times$6.0 mm$\times$5.0 mm is defined in which 400 spheres are densely packed. Consistently with the experimental setup, where the boundaries are smooth and rigid, we suppose that the NW-phase can invade along the side boundaries (the ``open-side'' condition).

The random packings are generated by DEM simulations. The PSD and porosity are defined as in the experiment. In order to reach the target value of porosity we employ a growth algorithm based on the REFD method (radius expansion-friction decrease) \cite{Chareyre2002}. This dynamic compaction method lets one control the porosity and it gives statistically homogeneous and isotropic microstructures. After this generation phase, the positions of the spheres are fixed. They don't move further during the drainage phase. 

For more generality, the data from experiment and simulations are all given in dimensionless forms hereafter.
The normalised capillary pressure $P^*$ is
\begin{equation}
 \label{equ:normalized}
 P^* = \frac{P^c \bar{D}} {\gamma^{nw}} 
\end{equation}
in which, $\bar{D}$ is the average size in the PSD and $\gamma^{nw}$ is the W-NW surface tension.

\subsection{Results and discussion}
\label{subsec:comparion_results_discussion}
\begin{figure}
 \centering
\includegraphics[width=0.6\columnwidth]{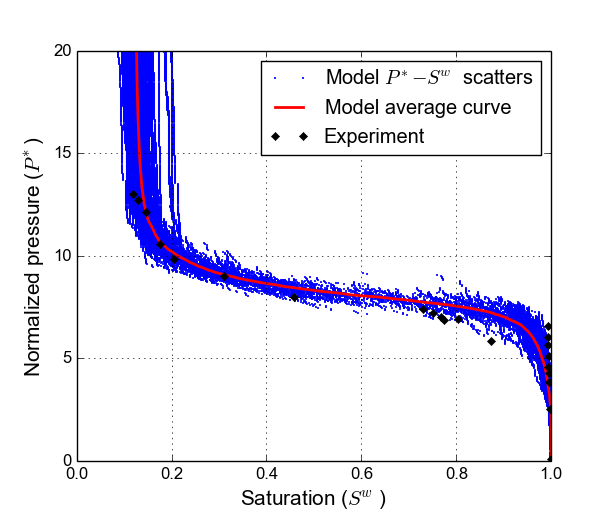}
\caption{Comparison between simulation and experiment for primary drainage $P^c-S^w$ curves. The number of observations of simulation is 100.}
\label{fig:vsExperimentNew}
\end{figure}

Using the technique described above, we compute the primary drainage process for 100 random packings having the same PSD and porosity.
Fig.\ref{fig:vsExperimentNew} presents the results of these simulations, in which we gather all scattered ($P^*,S^w$) points of each simulation in one image.
As shown in Fig.\ref{fig:vsExperimentNew}, although all packings share the same macro-scale parameters, the $P^*-S^w$ curves still have a distinct variability.
Especially, the residual saturation can differ significantly from one sample to another. 
This erratic dispersion could be reduced by enlarging the sample size. This trend will be be discussed in section \ref{sec:REV}. For the moment we keep the number of particles approximately equal to the number of particles in the scanned domain of the experiment.

The $P^*-S^w$ curves show a rather good agreement between the simulations and the experiments.  
The experimental data points are in the range of simulation ($P^*,S^w$) scatters, although the averaged curve differs slightly from the experimental one.
It can be explained by the simplifications done in the drainage model, by the fact that the real conditions are not well reflected in the boundary conditions of the subdomain or of the sample itself (cubic packing versus circular cross-section).

\begin{figure}
 \centering
\includegraphics[width=0.6\columnwidth]{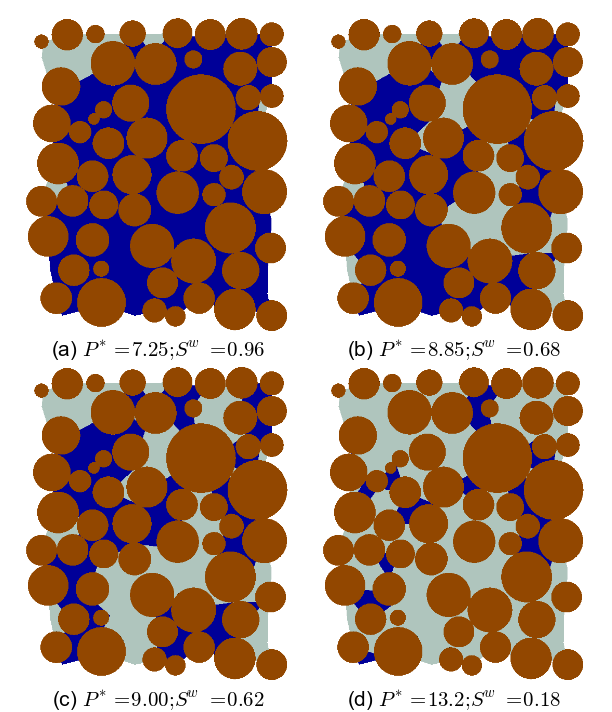}
\caption{The process of drainage (400 particles), NW-phase invade from top. Brown (gray) is solid phase, blue (black) is W-phase, and light cyan (white) is NW-phase, see colour version of this figure in the HTML.}
\label{fig:drainageProcess}
\end{figure}

We capture one test from the series of simulation and we cut a slice to observe the invasion patterns (Fig.\ref{fig:drainageProcess}).
When increasing $P^*$, the invasion starts from the pores with larger throat, in which the entry capillary pressure is smaller (see slice (a)). 
As described in previous section, the $P^c_e$ of pore throats is in average smaller along the boundaries than in the bulk, so that the NW-phase invades the side pores first (as shown in Fig.\ref{fig:drainageProcess}(a) and (b)).
Comparing slices (b) and (c) shows that under certain circumstances even a very small change in $P^*$ can cause a significant displacement of the NW-W interfaces.
For such event, Haines jumps go through large cluster of pores, causing a sharp decrease of W-phase content.
Slice (d) shows the end of the simulation when all the remaining W-phase is in the form of disconnected clusters entrapped by the NW-phase.

\begin{figure}
 \centering
\includegraphics[width=0.4\columnwidth]{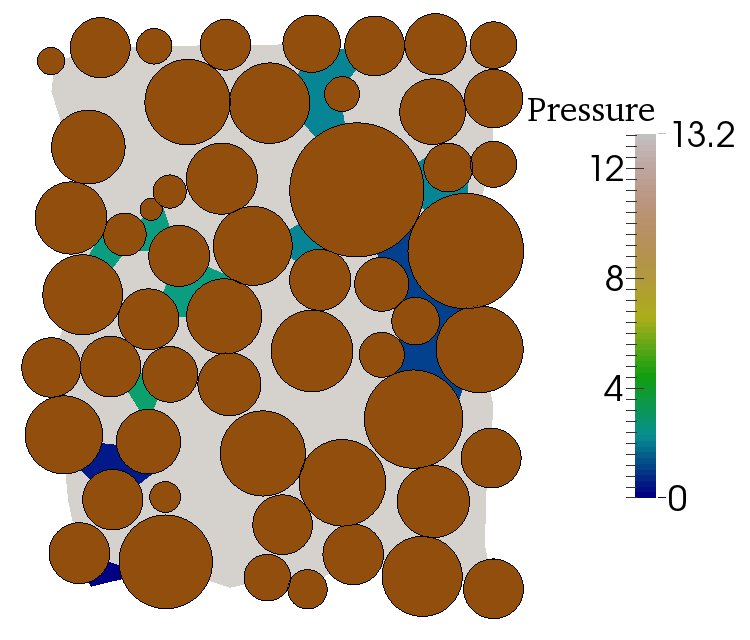}
\caption{Distribution of capillary pressure. Brown circle is solid phase, light region is NW-phase pressure $P^n$, dark regions are W-phase pressures $P^w$, see colour version of this figure in the HTML.}
\label{fig:pressureProfile}
\end{figure}

Since the model is neglecting W-phase transport by film flow or evaporation, these disconnected phases will never disappear. In a real situation, those processes would eventually lead to complete drainage.
As explained in section \ref{subsec:drainage_and_trapping}, the trapping sequences result in a different $P^w$ in each disconnected cluster. An example of this feature is shown in Fig.\ref{fig:pressureProfile}.
From the pressure distribution we can determine the order of the disconnections, with the larger $P^w$ corresponding to earlier entrapment.

\section{Representative element volume (REV)}
\label{sec:REV}
In practice one never deals with infinite systems and it is necessary to understand the effects of sample size, sample shape, and gradients of state variables. This difficulty exists in both experiments and numerical simulations of - so called - representative element volume (REV). It is amplified in the context of microscale experiments and simulations since they tend to study domains of smaller size compared to conventional lab or field experiments. REV often refers to a sample size sufficiently large to provide statistical robustness to an averaging procedure \cite{Bear1972}. 
This definition is rather clear and its application is straightforward if the variable of interest is the average porosity of a statistically homogeneous material, for instance. In such case statistics generated by sub-sampling a large domain are enough to determine variance reduction as a function of size: the variance $\sigma^2$ decreases as $\alpha/V$ as soon as $V\gg \alpha$, where $\alpha$ depends on the size of the heterogeneities and $V$ is the sampled volume ($\sigma$ is the standard deviation). Defining the volume by the number of spheres it contains leads to an equivalent proportionality between $\sigma$ and $1/\sqrt{N}$. Knowing $\alpha$, the REV size depends only on the decision of which variance it tolerated for a single measurement.
Moreover, the excessive variability of results on small samples can always be mitigated by averaging the results on many samples.

It is very important to note that in the above context the minimal REV size is not a fixed value. It depends strongly on the tolerated deviation (a tolerated standard deviation decreased by 10 results in REV volume multiplied by 100). This is overlooked in many papers in which some differences are said ``negligible'' without a clear definition of how small ``negligible'' is, which makes the determination of $\alpha$ impossible.

Two phase flow (among other processes) adds complexity to the problem in a way which is not always very well acknowledged. The question is not only to control the scattering of results, but also to make sure that the drainage process itself is not influenced by the size (and shape) of the sample. In other words there is a need to know if, for a particular $P_c$ imposed on samples, the saturation will be simply distributed around a unique size-independent mean value, or if the mean value itself can be biased by the sample size.    
In this section we examine both aspects, {\it i.e.}, 1) how samples of different sizes result in different values in average and 2) how the results on sub-samples fluctuate around the mean. The effect of sample shape is analysed in the last part.
%
%

\subsection{Sample size}
We report a series of simulated tests with different sizes of cubic samples. 
The number of spheres ($N$) ranges from 100 to 40,000, and for every $N$ the simulations are repeated on 100 different sphere assemblies. For a given $N$ the samples only differ in the positions of individual spheres. Porosity and PSD are the same as in the previous section.
Both open-side and closed-side modes are considered in the tests for comparisons.
The same averaging technique as shown in sec.\ref{sec:comparison_with_experiment} is used to manipulate the statistical results, {\it i.e.}, for each size of sample, an averaged $P^*-S^w$ curve is achieved based on the 100 observations. For each $P^*$ value the standard deviation $\sigma(S^w)$ of saturation is calculated.

\begin{figure}
 \centering
\includegraphics[width=0.6\columnwidth]{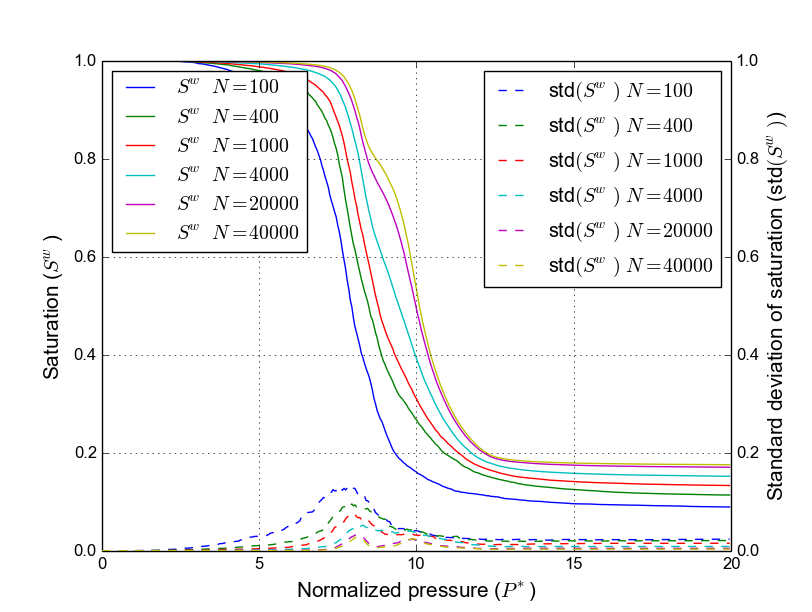}
\caption{Average $P^*-S^w$ curve as a function of sample size in open-side mode (100 repeated simulations for each case). See colour version of this figure online.}
\label{fig:openRev}
\end{figure}

\begin{figure}
 \centering
\includegraphics[width=0.6\columnwidth]{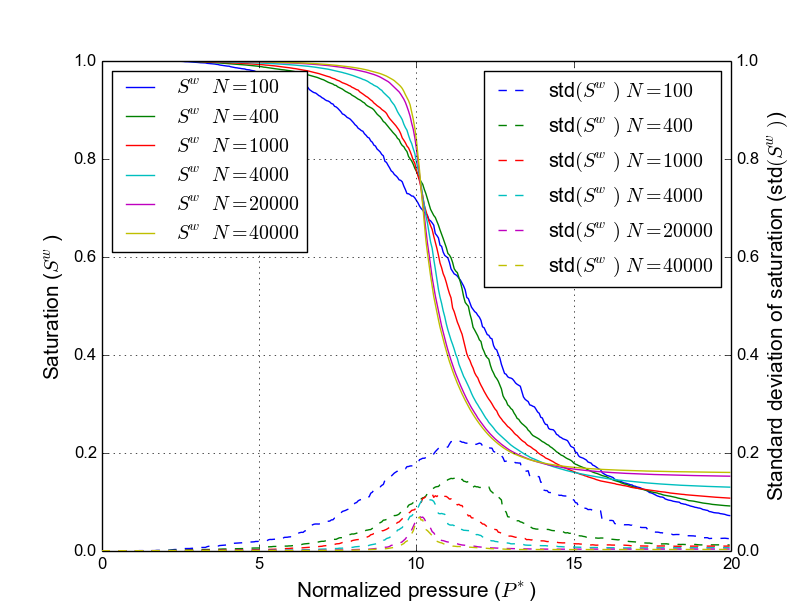}
\caption{Average $P^*-S^w$ curve as a function of sample size in closed-side mode (100 repeated simulations for each case). See colour version of this figure online.}
\label{fig:closedRev}
\end{figure}

As seen in Fig.\ref{fig:openRev} and Fig.\ref{fig:closedRev}, the averaged $P^*-S^w$ curves for different sizes are clearly distinct.
In open-side drainage, the shift of the $P^*-S^w$ curve with $N$ is monotonic.
Under the same $P^*$, a larger $N$ results in a larger $S^w$.
In closed-side mode, the effect of $N$ is more complex. The curves are not simply shifted as they intersect each other.
Drainage starts for smaller $P^*$ values in small samples. There is a transition near $P^*=10$ (corresponding to $S^w\simeq 0.7$), after which the ordering of the curves is reverted and small samples have larger degrees of saturation. A second inversion occurs before reaching the residual saturation ($P^*\simeq16$).
The shapes of the curves are clearly different between Fig.\ref{fig:openRev} and Fig.\ref{fig:closedRev} for $0.9>S^w>0.2$. 
In open-side drainage, all curves have very similar slopes while in closed-side drainage smaller samples have more shallow slopes.

In both drainage modes larger samples have larger residual saturation. A result which might be explained by the possibility to form clusters of trapped W-phase of larger sizes in larger samples (it may also explain the shift of the curves in Fig.\ref{fig:openRev}).

\begin{figure}
 \centering
\includegraphics[width=0.6\columnwidth]{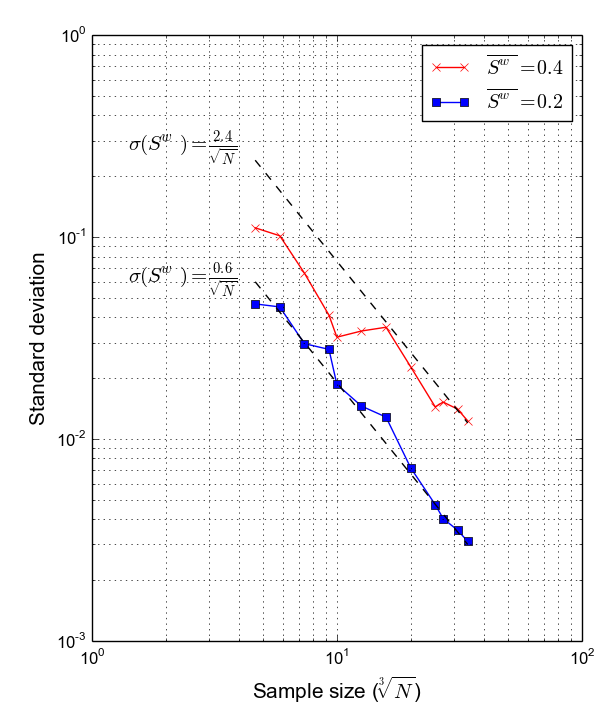}
\caption{The relationship between standard deviation of saturation and packing size (in open-side drainage mode).}
\label{fig:rev1SwStdLog}
\end{figure}

In Fig.\ref{fig:openRev} and Fig.\ref{fig:closedRev}, the $\sigma(S^w) - N$ curves show how larger samples narrow the distribution of $S^w$ on different samples.
The peaks of $\sigma(S^w)$ correspond to the major evolution of $S^w$, when small changes of $P^*$ lead to the recursive invasion of many pores. This dispersion is much smaller in open-side mode.
The decreasing trend of $\sigma(S^w)$ and $N$ in open-side mode is illustrated for selected values of saturation, ${S^w}=0.4$ and ${S^w}=0.2$.
The $\sigma(S^w) - \sqrt[3]{N}$ curves are reported on Log-Log axes in Fig.\ref{fig:rev1SwStdLog}. $\sqrt[3]{N}$ can be interpreted as the edge length of a cubic domain containing $N$ spheres. In the figure, fitting equations following conventional variance reduction are superimposed. They agree with the data in a satisfying manner for the larger sizes.
\begin{equation}
 \label{equ:rev1std40}
 \sigma(S^w)=\frac{0.6}{\sqrt{N}}
\end{equation}
when $\overline{S^w}=0.2$; and
\begin{equation}
 \label{equ:rev1std20}
 \sigma(S^w)=\frac{2.4}{\sqrt{N}}
\end{equation}
when $\overline{S^w}=0.4$.

Regardless of the drainage mode at the boundaries, it can be concluded that the average $S^w$ is significantly biased for $N<20,000$, and the bias is significantly larger than the standard deviation. For instance, the standard deviation of residual saturation for $N=100$ is of the order of 0.02 (0.13 when $S^w=0.4$) while the difference with $N=40000$ in terms of the average residual saturation  is about 0.1 (0.34 for the value of $P^*$ corresponding to $S^w=0.4$ in the large sample). The difference in $S^w$ between $N=20,000$ and $N=40,000$ is 0.002 at residual saturation, and 0.02 near $S^w=0.4$.
Open-side conditions seem to give more robust measurements. They preserve the general shape of the $P^*-S^w$ curves for every $N$ and the standard deviation is decreased.
Interestingly, it suggests that the simulations compared well with the experiments (previous section) only because they were both biased in the same way ($N\approx400$ spheres in both cases), while simulating larger samples would have led to a worse agreement.

\subsection{Subsampling}
In order to examine the sub-sampling problem, we used a large sample of 64,000 spheres in which we defined a set of points (64 vertices of a cubic $4 \times 4 \times 4$ array) to be the centers of subdomains. The porosity and saturation per subdomain are analyzed for different sizes of the subdomains. The PSD and average porosity are the same as in previous sections.
Like before the subdomain size is defined by $\sqrt[3]{N}$ where $N$ is the number of sphere per subdomain (in average). 
Based on the conclusion of previous section, we examine only the open-side drainage mode.

\begin{figure}
 \centering
\includegraphics[width=0.6\columnwidth]{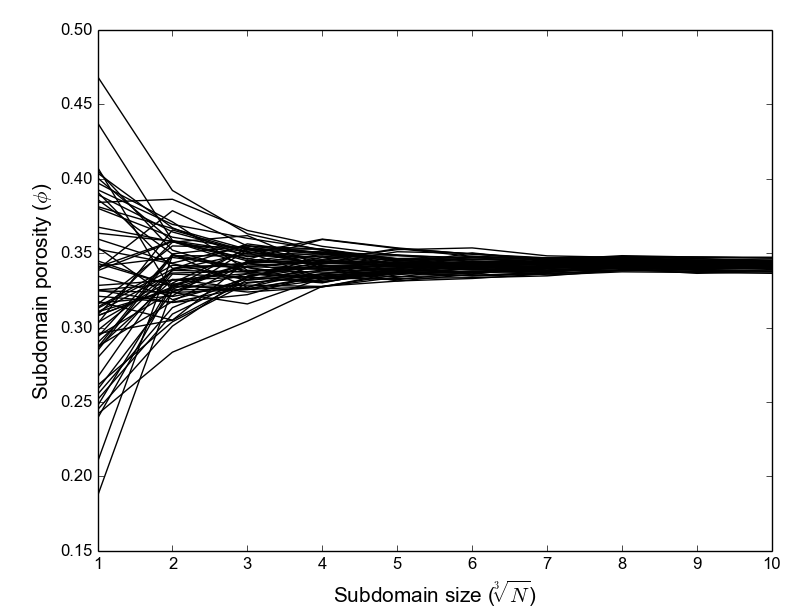}
\caption{Micro-porosity versus averaging subdomain size. $\overline{\phi_{sub}}=0.34$ }
\label{fig:porosity}
\end{figure}

\begin{figure}
 \centering
\includegraphics[width=0.6\columnwidth]{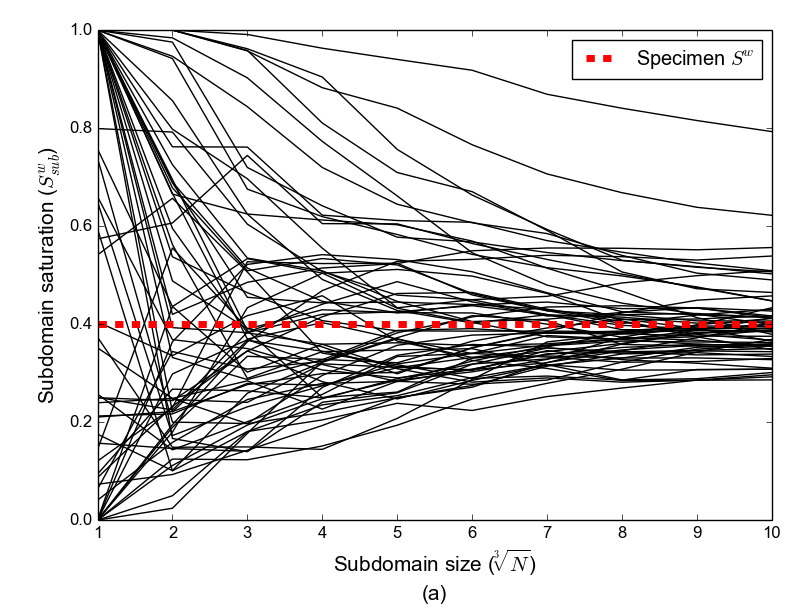}
\includegraphics[width=0.6\columnwidth]{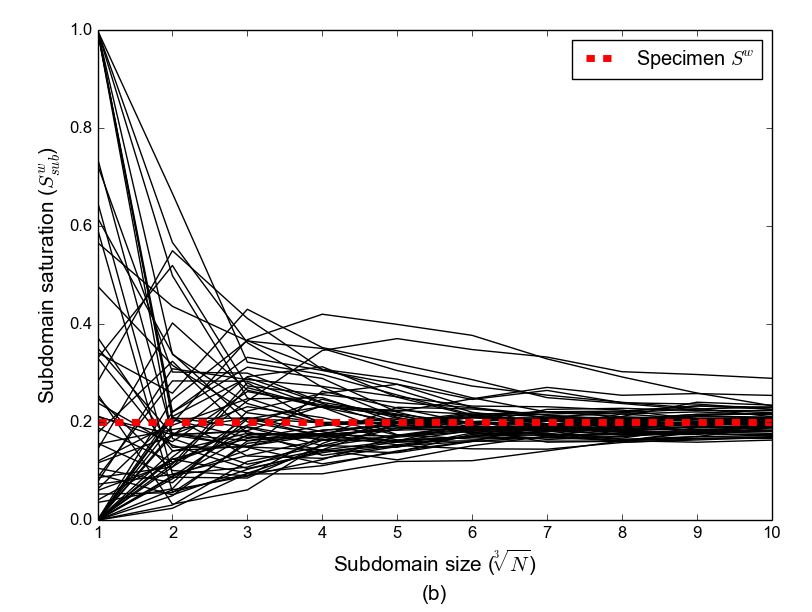}
\caption{Micro-saturation versus averaging subdomain size. (a): $\overline{S^w_{sub}}=0.4$, (b): $\overline{S^w_{sub}}=0.2$. }
\label{fig:revSw}
\end{figure}

The average quantities obtained in each subdomain are plotted as functions of $\sqrt[3]{N}$ and superimposed in Fig.\ref{fig:porosity} and Fig.\ref{fig:revSw}. As expected, the porosity of every subdomain converges steadily to the global porosity of the sample as $N$ increases. The evolution of the $S^w_{sub}$ values is more erratic. It shows strong oscillations for the smaller sizes, much more scattering than porosity, and some subdomains hardly converge to a general trend even for $10^3$ spheres. Obviously this can be explained by large single-phase clusters. 

\begin{figure}
 \centering
\includegraphics[width=0.6\columnwidth]{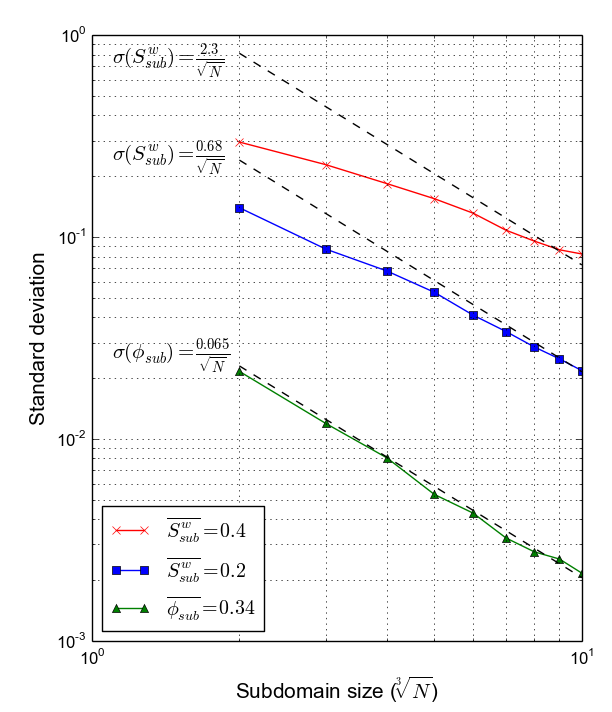}
\caption{Statistics of saturation and porosity versus size of the averaging subdomain.}
\label{fig:rev2SwStdLog}
\end{figure}

Fig.\ref{fig:rev2SwStdLog} reports the evolution of $\sigma(\phi_{sub})$ and $\sigma(S^w_{sub})$ as functions of $\sqrt[3]{N}$.
The fact that $\sigma(\phi_{sub}) < \sigma(S^w_{sub})$ implies that the minimal REV size for saturation is clearly much larger than what could be used for estimating porosity. A conclusion also reached by \cite{Hilpert2001}, in which standard deviations of similar magnitude are reported.
Fig.\ref{fig:rev2SwStdLog} also shows trend lines of the form $\alpha / {\sqrt{N}}$, where $\alpha$ is adjusted for each series of points. The $\sigma(S^w_{sub}) - \sqrt[3]{N}$ is very well described by
\begin{equation}
 \label{equ:rev2stdPorosity}
 \sigma(\phi_{sub})=\frac{0.065}{\sqrt{N}}.
\end{equation}
The evolution of $\sigma(S^w_{sub})$ does not follow such a simple form. Trying to adjust $\alpha$ using the data from the largest samples (fig.\ref{fig:rev2SwStdLog}) suggests
\begin{equation}
 \label{equ:rev2std40}
 \sigma(S^w_{sub})=\frac{2.3}{\sqrt{N}}
\end{equation}
for $\overline{S^w_{sub}}=0.4$, and 
\begin{equation}
 \label{equ:rev2std20}
 \sigma(S^w_{sub})=\frac{0.68}{\sqrt{N}}
\end{equation}
when $\overline{S^w_{sub}}=0.2$.
The expressions are very close to the ones found in previous section.

\begin{figure}
 \centering
\includegraphics[width=0.6\columnwidth]{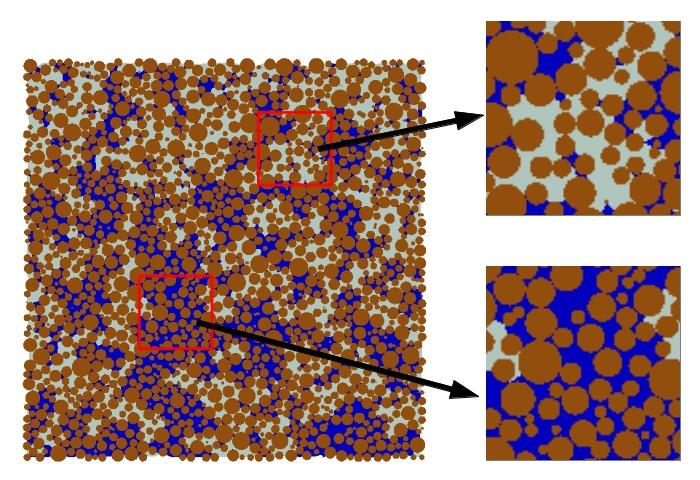}
\caption{Right: a slice through a sample at ${S^w_{sub}}=0.4$ (the top boundary is connected to the NW reservoir), in which strong capillary fingering can be observed. Left: zoom on two subdomains  used in the averaging procedure.}
\label{fig:sw40slicezoom}
\end{figure}

The large differences between those standard deviations are easily explained.
At higher saturation ($\overline{S^w_{sub}}=0.4$), capillary fingerings results in large single-phase patches (see Fig.\ref{fig:sw40slicezoom}). Hence the average saturation computed in a particular subdomain is strongly influenced by the position of its center. If the subdomain is too small, it may even be entirely occupied by one of the phases cluster. The condition that the sampled volume must be larger than the heterogeneities is never satisfied, which leads to the poor agreement with the $1/\sqrt{N}$ trend. This does not happen for porosity, since every solid particle is surrounded by a certain amount of pore space (especially for spherical shapes). 
%

\begin{figure}
 \centering
\includegraphics[width=0.6\columnwidth]{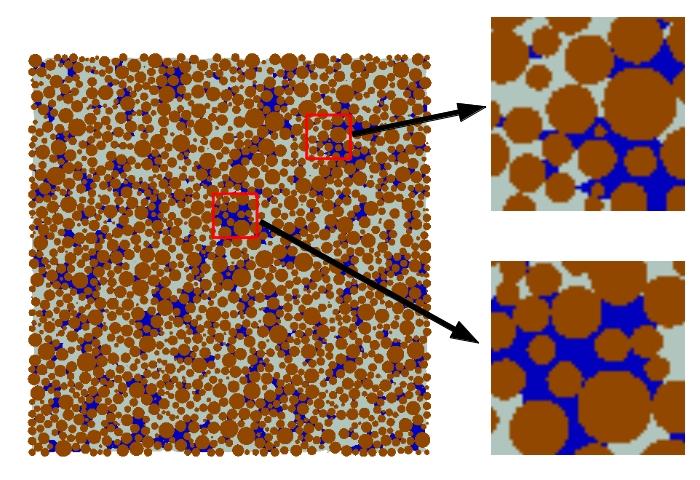}
\caption{Right: a slice through a sample at ${S^w_{sub}}=0.2$ (the top boundary is connected to the NW reservoir), in which patches of trapped W-phase are visible. Left: zoom on two subdomains  used in the averaging procedure.}
\label{fig:sw20slicezoom}
\end{figure}

At lower saturation ($\overline{S^w_{sub}}=0.2$) the receding W-phase is present mainly in the form disconnected patches (Fig.\ref{fig:sw20slicezoom}).
These patches are larger than a particle diameter but smaller than the patches observed at $\overline{S^w_{sub}}=0.4$. The standard deviation is clearly reduced and the $1/\sqrt{N}$ trend is nearly acceptable for the largest samples. The $\alpha$-values which appear in the fitting equations (0.065 for $\sigma(\phi_{sub})$ versus 0.68 for $\sigma(S^w_{sub})$ for $\overline{S^w_{sub}}=0.2$) suggest that the heterogeneities of the phase distribution have a characteristic volume one hundred times larger than the heterogeneities of the void space. A result consistent with the image of Fig.\ref{fig:sw20slicezoom} where we may accept 5 particle diameters ($5^3\simeq100$) to reflect the typical distance separating the disconnected patches.
The poor fit obtained with $1/\sqrt{N}$ evolution suggests that even the largest subsample (1000 spheres) is far below an acceptable REV size when $\overline{S^w_{sub}}>0.4$. We did not proceed to larger sizes since the subdomains would overlap each other or reach the boundaries. The strong size dependency of saturation near the percolation threshold (the value of $P^*$ for which the NW-phase reaches the W-reservoir) is actually a known issue: the size of the largest patch tend to increase as the sample size is increased \cite{Degennes1978,Lenormand1980,Wilkinson1983}, hence the REV question in itself is ill-posed. This is a rather challenging problem for defining macroscale properties. We shall not enter this debate here as it would need more investigation (possibly using our model). The reader may refer to \cite{Hilpert2001} for a fractal approach of the problem. Here we retain that the standard deviation seems to reach a normal trend at least at low saturation.   

\section{Boundary conditions and aspect ratio}
\label{sec:side_boundary_size_effect}
In this section, we analyse the effects of the side boundary conditions and the sample shape. 
Since circular columns are out of reach of our current algorithm we only discuss the shape effect in terms of the aspect ratio of rectangular boxes.
The samples are made of 40000 spheres with the same porosity and PSD as before.
They are prepared with different ratio of cross-sectional side-length $L$ over height $H$ ($H/L=$0.5, 1.0 and 5.0).

\begin{figure}
 \centering
\includegraphics[width=0.4\columnwidth]{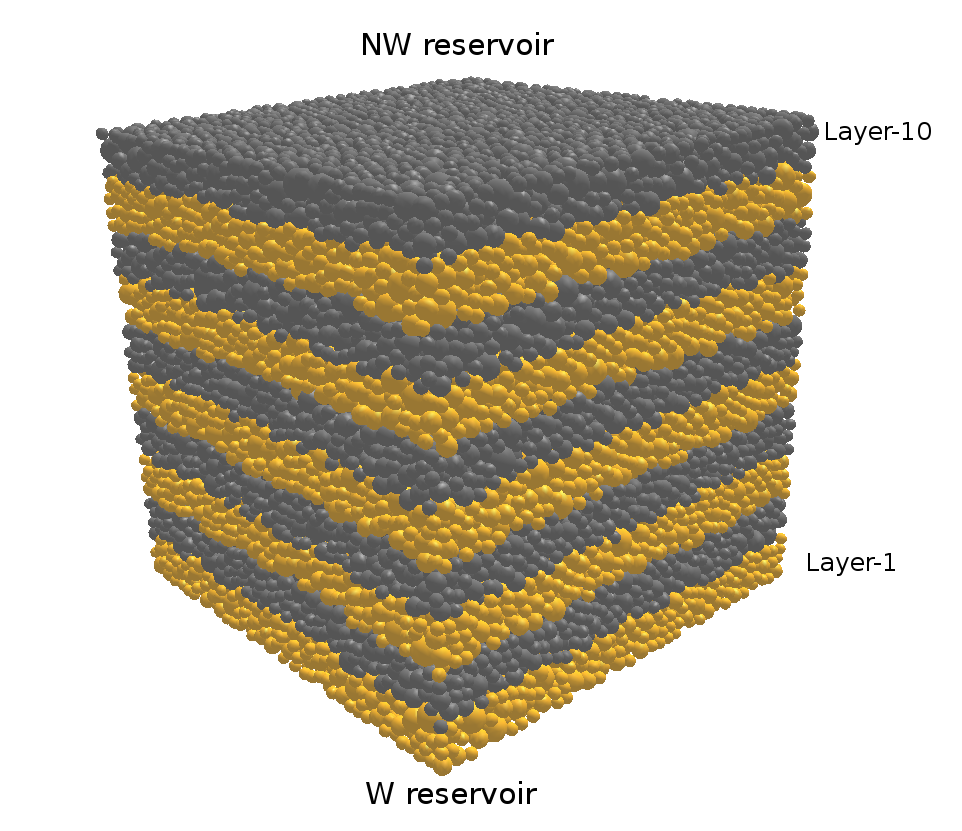}
\caption{Numerical setup and indexed layers.The bottom of Layer-01 is connected to the W-phase reservoir, the top of Layer-10 is connected to the NW-phase reservoir.}
\label{fig:sampleLayers}
\end{figure}

A sub-sampling is done by dividing the column in 10 layers perpendicular to the drainage direction.
These layers are indexed from ID-1 (connected to W-phase reservoir) to ID-10 (connected to NW-phase reservoir), as shown in Fig.\ref{fig:sampleLayers}.
Since the random positioning of spheres can influence the results, as illustrated in section \ref{subsec:comparion_results_discussion} (see Fig.\ref{fig:vsExperimentNew}), we report results averaged in each layer for different samples.
In order to describe the invasion of the NW-phase, we define the NW-phase penetration depth $D^p$ as the maximum vertical distance between the NW-reservoir and the NW-W interface.
This penetration depth is normalised as
\begin{equation}
 \label{equ:normalized_depth}
 D^* = \frac{D^p} {H} 
\end{equation}
in which $H$ is the height of specimen.
Both open-side and closed-side drainage modes are simulated. 
The $P^*$ and $S^w$ values are recorded for the entire packing and per layer. 

\subsection{Effect of side boundary connectivity}
\label{subsec:effecet_of_side_boundary}
\begin{figure}
 \centering
\includegraphics[width=1.\columnwidth]{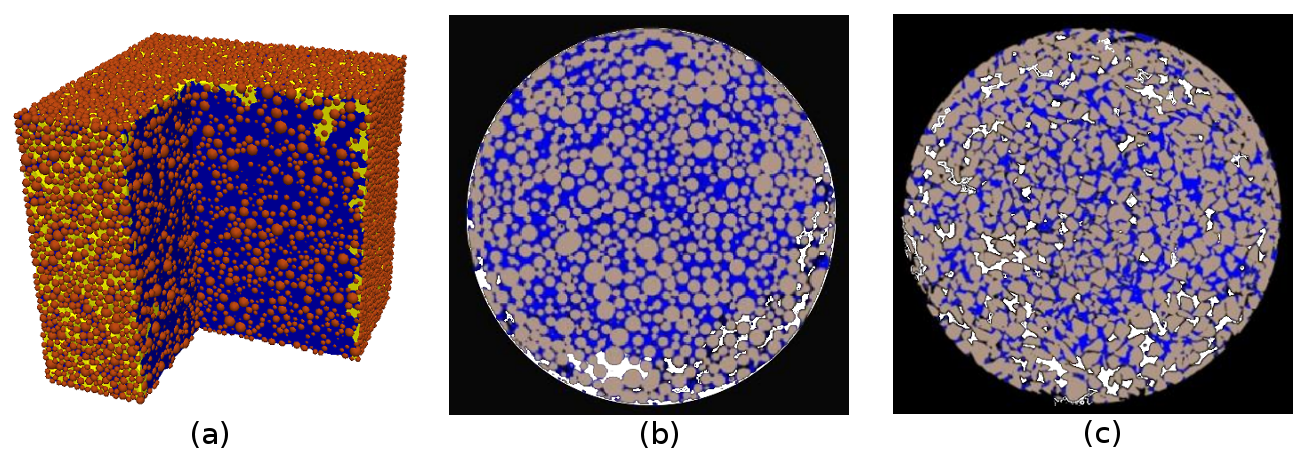}
\caption{Comparison of invasion between a simulation in open-side drainage and the experiments of Khaddour {\it et al.} \cite{Khaddour2013}. (a) Simulation (dark blue: W-phase, yellow: NW-phase). (b) experiment on glass beads. (c) experiment on Hostun sand (gray: glass beads/sand, dark blue: water, white: air).}
\label{fig:scansGhonwa}
\end{figure}

In open-side drainage, the NW-phase can invade all pores, including the pores in contact with the side boundaries of the container. In closed-side conditions, the invasion is only allowed in the inner part of the sample.
Examining early stages of the the open side drainage reveals a preferential invasion starting along the boundaries (Fig.\ref{fig:scansGhonwa}(a)). This tendency has been also observed in experiments on glass beads \cite{Khaddour2013}, as shown in Fig.\ref{fig:scansGhonwa}(b). Simply, the throats formed by two spheres in contact with a flat surfaces tend to be larger than the throats found in the rest of the microstructure ({\it i.e.}, between 3 spheres). It is consistent with previous findings on anomalous porosity due to wall effects \cite{Marketos2010} and it leads to lower values of $P^c_e$ along the boundaries, hence preferential invasion. In a second step the invading phase percolate to the inner part starting from all boundaries of the samples (W-reservoir excepted).

It is worth noting that this drainage sequence may not be generalised to every granular material since the experiments in \cite{Khaddour2013} did not show the same evolution with grains of irregular shapes (Fig.\ref{fig:scansGhonwa}(c), Hostun sand). Angular or elongated grains thus seem less prone to form large throats near the walls of the container.

The effect of side boundary conditions can be further investigated by comparing the invasions under different assumptions.
At the beginning of invasion, the layer saturation is very heterogeneous, with low saturation near the NW reservoir and high saturation near the W reservoir.
This phenomenon can be observed in both drainage modes as shown in Fig.\ref{fig:boundTrueLayers} ($P^*=8.0$) and Fig.\ref{fig:boundFalseLayers} ($P^*=9.0$), where saturation decreases almost only in the top layer.
This effect is less pronounced in open-side mode.
By comparing the distribution of $S^w$ for the same aspect ratio of specimens, {\it i.e.}, Fig.\ref{fig:boundTrueLayers}(a) {\it vs} Fig.\ref{fig:boundFalseLayers}(a), Fig.\ref{fig:boundTrueLayers}(b) {\it vs} Fig.\ref{fig:boundFalseLayers}(b), and Fig.\ref{fig:boundTrueLayers}(c) {\it vs} Fig.\ref{fig:boundFalseLayers}(c), it is found that the W-phase retention is much more homogeneous if the side boundaries can be invaded, as could be expected from the aforementioned two-step sequence of invasion.
In closed-side mode, the NW-phase can only invade layer by layer, leading to a stronger saturation gradient in intermediate steps.
Finally, the residual saturation is approximately homogeneously distributed in all layers.

The evolution of saturation and penetration with capillary pressure are shown in Fig.\ref{fig:boundTruePcSwDp} and Fig.\ref{fig:boundFalsePcSwDp}.
It is found that the NW-phase invasion in open-side mode starts at lower values of $P^*$, this is due to large pores along the boundaries. 
The evolution of $D^*$ confirms the two-step sequence in open-side drainage: the main evolution of saturation happens after NW-phase percolation through the entire sample, {\it i.e.}, after $D^*=1.0$. With closed-side drainage the main evolution of $S^w$ is accompanied by the increase of $D^*$.
Again, the results are less scattered when the side boundaries can be invaded.

\subsection{Effect of aspect ratio}
\label{subsec:effect_of_aspect_ratio}
\begin{figure}
 \centering
\includegraphics[width=0.8\columnwidth]{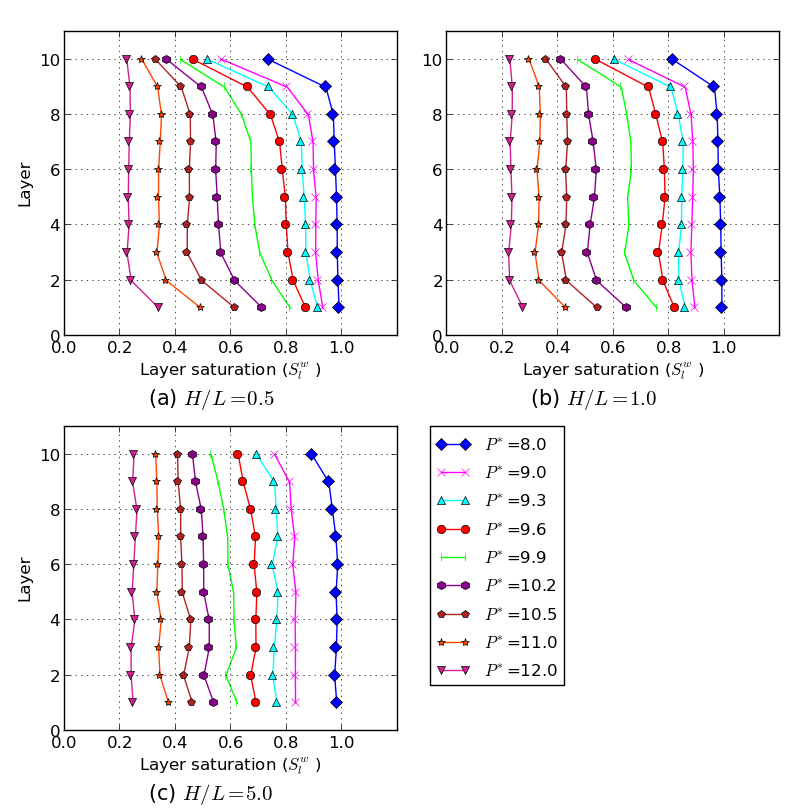}
\caption{Saturation distribution of different layers under certain capillary pressures in open-side drainage.}
\label{fig:boundTrueLayers}
\end{figure}

\begin{figure}
 \centering
\includegraphics[width=0.8\columnwidth]{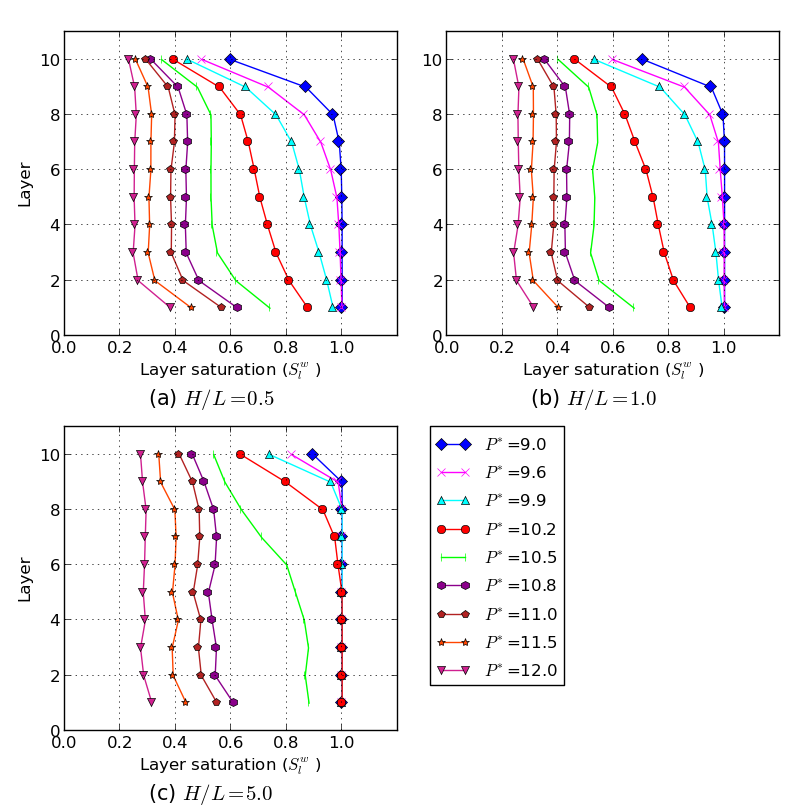}
\caption{Saturation distribution of different layers under certain capillary pressures in closed-side drainage.}
\label{fig:boundFalseLayers}
\end{figure}

\begin{figure}
 \centering
\includegraphics[width=0.6\columnwidth]{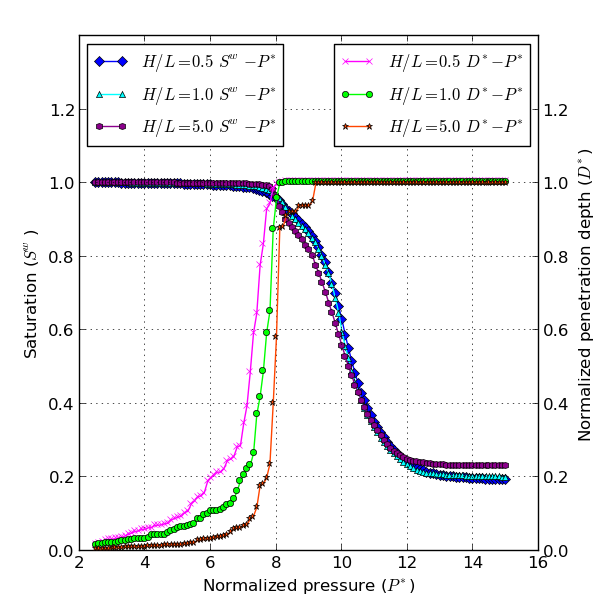}
\caption{Comparison between different aspect ratios of simulations for $S^w-P^*$ curves and $D^*-P^*$ curves in open-side drainage}
\label{fig:boundTruePcSwDp}
\end{figure}

\begin{figure}
 \centering
\includegraphics[width=0.6\columnwidth]{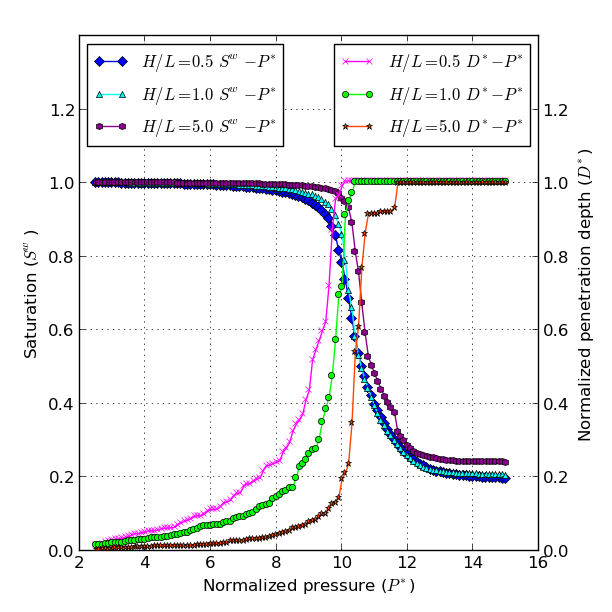}
\caption{Comparison between different aspect ratios of simulations for $S^w-P^*$ curves and $D^*-P^*$ curves in closed-side drainage}
\label{fig:boundFalsePcSwDp}
\end{figure}

In Fig.\ref{fig:boundTruePcSwDp} or Fig.\ref{fig:boundFalsePcSwDp}, for a given $P^*$, the NW-phase invade more deeply for smaller $H/L$ ratio.
This is consistent with the W-phase profiles of Fig.\ref{fig:boundTrueLayers} and Fig.\ref{fig:boundFalseLayers}, which show a lower saturation for smaller aspect ratio.
This applies equally well to residual saturation, which suggests more W-phase trapping for larger $H/L$.

This effect is less significant in open-side drainage for the layers 3-7 (Fig.\ref{fig:boundTrueLayers}).
In this case most of the difference in sample saturation comes from those layers within a short distance from the reservoirs. A large $H/L$ tends to reduce the fraction of the total volume which is exposed to this near-reservoir situation, hence for $H/L=5$ even layers 1 and 10 only slightly deviate from the global average.   

In closed-side drainage, on the other hand, a smaller $H/L$ tends to produce slightly more homogeneous phase distribution - even though it remains rather heterogeneous for $0.4<S_r<1$ (Fig.\ref{fig:boundFalseLayers}). This evolution is dominated by a main percolation event, corresponding to large gradients of saturation. After the first percolation ({\it i.e.}, as soon as $D^*=1$) a progressive homogenisation of the phases distribution occurs until residual saturation is reached. As percolation occurs a bit earlier at low $H/L$, the homogenisation phase starts earlier too. In any case, closed-side boundary conditions do not provide a robust base for evaluating the $P^*-S^w$ relation for $S^w>0.4$ in primary drainage, given the large heterogeneity of the phase distribution.

\section{Conclusion}
\label{sec:conclusion}
A pore-scale model of quasi-static two-phase flow in dense packings of poly-disperse spheres has been proposed.
In the proposed decomposition the pore throats are planar objects defined by the facets of the triangulation, while all the pore space is contained in the volume of the pore bodies. Drainage occurs by a recursive invasion of the pores when the capillary pressure exceeds threshold (``entry'') values defined locally for each pore throat. The model captures the entrapment of the receding W-phase, resulting in a residual saturation. The simulations of primary drainage are in good agreement with experiments in terms of $P^*-S^w$ relation and also with regard to the preferential invasion along the boundaries.

With our current implementation the complete network generation is achieved in less than 10s for 64,000 spheres on a standard workstation (Intel Xeon 2.80GHz, executed on one single core); the cost of an invasion step (finding the stable phase distribution corresponding to one value of $P^c$) depends on the number of invaded pores hence fluctuates during a simulated drainage, for 64,000 spheres it never exceeds 0.1s. Further work is needed to implement arbitrary sample shapes, this will let us match experimental setups (circular cross sections in most cases) more precisely. The extension to periodic boundary conditions is also a work in progress.

One goal of this work was to assess size effects and boundary effects on primary drainage when testing small samples - a key question when designing small scale experiments and simulations.
We examined separately the statistics from samples of different sizes, then those from subsamples of a single large sample. The main conclusions are:
\begin{itemize}
 \item The standard deviation of $S^w$ in repeated simulations of primary drainage roughly follow a simple variance reduction law with increasing sample size, {\it i.e.}, $\sigma(S^w)\simeq 2.4/\sqrt{N}$ for the maximum deviation (when $S^w$ is close to 0.4) and $\sigma(S^w)\simeq 0.6/\sqrt{N}$ at residual saturation. The orders of magnitude of $\sigma(S^w)$ are in agreement with \cite{Hilpert2001} in which 2500 was suggested as a sufficient number of spheres, corresponding to a maximum deviation of $\sigma(S^w)=0.048$ with our expression.
 \item Standard deviation should not be the unique criterion for evaluating the representativity. Indeed the sample size can be the cause of significant bias in the average result. The saturation decreases with decreasing sample size. For 2500 spheres for instance, the difference may reach 0.3 based on our results (for $S^w\simeq0.4$), {\it i.e.}, much more than the standard deviation found for this particular size. We found that at least 20000 spheres must used in order to reduce the bias below 0.02.
 \item Boundary conditions also affect the result significantly. The paradox is that the strong boundary effects observed when preferential invasion occurs along the boundaries lead to more robust evaluations of the $P^*-S^w$ relation overall. When this phenomenon is not present the shape of the $P^*-S^w$ curve is more sensitive to sample size and the phase distribution always show strong gradients of saturation. A similar problem would most likely appear with periodic boundary conditions.
\end{itemize}
The need to compute large samples clearly shows the need for efficient numerical techniques such as the pore-network methods. It justifies {\it a posteriori} our attempt to develop a fast pore-scale method for coupled hydromechanical problems.

We suggest a few guidelines for further attempts to compare simulations and small scale experiments:
\begin{itemize}
 \item The comparisons should be done on samples of similar sizes, ideally similar shapes, and with the same boundary conditions. Even below the REV size, this can lead to relevant model validations provided that the inherent variability is kept in mind.
 \item The experiments should be designed and reported in such a way that the boundary conditions can be accurately reproduced in a model. Scanning a small window in a long column is detrimental for this reason.
 \item Ideally, the position and size of each grain should be provided to eliminate the main source of variability, this is within reach of recent techniques \cite{Ando2012}.
\end{itemize} 

The above conclusions apply to well controlled small-sized specimen of granular material with statistically homogeneous distribution of porosity. The extrapolation to conventional lab or field experiments should be done with care. Preferential boundary invasion may occur in some real tests but it could be a peculiar feature of spherical (or well-rounded) grains only. If this boundary invasion is not present, then the lab tests may be more similar to our closed-boundary case, which suggest the occurrence of strong gradients of saturation in the samples. It raises difficult questions on the interpretation of lab tests: is an average $S^w$ relevant when strong gradients of $S^w$ are present?
More generally, our results underline a known feature: the key role of heterogeneities (in our case the boundaries) in the drainage process. Every heterogeneity of a soil sample (be it intrinsic or due to a particular sampling technique) may strongly influence its water retention properties. Likewise, heterogeneities at the field scale (soil composition, roots, wormholes,...) may play a dominant role in the transfers.

\section{Acknowledgements}
We thank G. Khaddour and S. Salager for discussions and for providing us with the tomography images (Fig.\ref{fig:scansGhonwa} b and c), G. Viggiani for inspiring this cooperation, and E. Ando for proofreading the manuscript.
The first author acknowledges support by the China Scholarship Council.

\section{Appendix}
\subsection{Calculation of capillary force and tension force for a pore throat}
\begin{figure}
 \centering
\includegraphics[width=0.6\columnwidth]{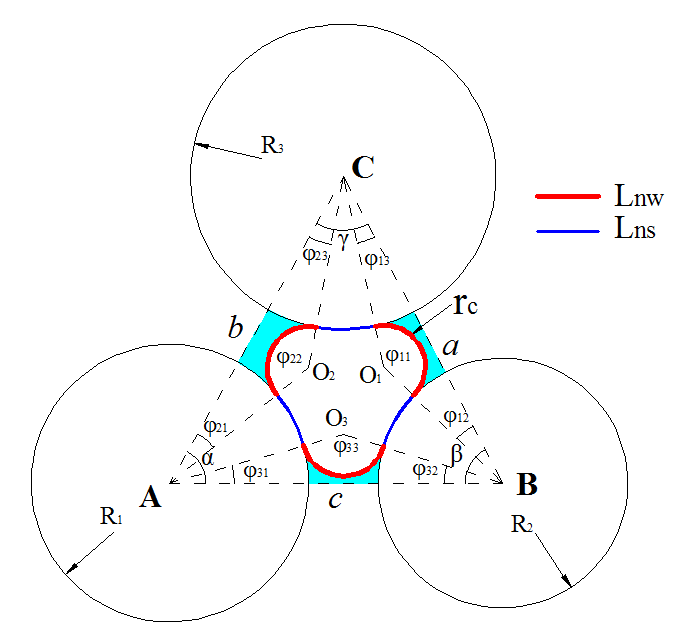}
\caption{Cross section geometry of pore throat, the pore throat radius $r_c$ is determined based on the balance of forces for non-wetting/wetting interface. }
\label{fig:poreThroatGeometry}
\end{figure}

In this appendix, we give explicit expressions of the capillary force $F^p$ and tension force $T^{\gamma}$ acting at a pore throat.

We consider a pore throat (see Fig.\ref{fig:poreThroatGeometry}) defined by the radii and positions of neighboring solid particles.
For a particular radius of curvature $r_c$ of the NW interfaces and in perfectly wetting condition (contact angle $\theta=0$) the contact line between NW and W phases is tangent to the solid surface. 

The area of the triangle $\Delta ABC$ can be written as follows:
\begin{equation}
 \label{equ:Area_ABC}
 A _{\Delta ABC} = \frac{1}{2}bc\sin \alpha
\end{equation}
Using laws of cosines, we can write the following equations to solve $\alpha$, $\beta$ and $\gamma$ in $\Delta ABC$,
\begin{equation}
 \label{equ:a^2}
 a^2 = b^2 +c^2 - 2bc\cos \alpha
\end{equation}
\begin{equation}
 \label{equ:b^2}
 b^2 = a^2 + c^2 - 2ac\cos \beta
\end{equation}
\begin{equation}
 \label{equ:c^2}
 c^2 = a^2 +b^2 -2ab\cos \gamma
\end{equation}
Likewise, the areas and $\varphi_{ij}$ in $\Delta AO_3B$, $\Delta BO_1C$ and $\Delta AO_2C$ can be obtained. 

The total area of liquid bridge $A_{lb}$ within the throat's section (the part occupied by the W-phase) is
\begin{equation}
\begin{split}
 \label{equ:liquid_bridge_area}
 A_{lb} = (A_{\Delta AO_3B} - 0.5R_1^2 \varphi _{31} - 0.5R_2^2 \varphi _{32} - 0.5r_c^2 \varphi _{33}) 
          \\+ (A_{\Delta BO_1C} - 0.5R_2^2 \varphi _{12} - 0.5R_3^2 \varphi _{13} - 0.5r_c^2 \varphi _{11})
          \\+ (A_{\Delta AO_2C} - 0.5R_1^2 \varphi _{21} - 0.5R_3^2 \varphi _{23} - 0.5r_c^2 \varphi _{22})
\end{split}
\end{equation}
The area $A_n$ corresponding to the invading NW-phase, and on which $p_c$ is exerted can finally be evaluated as:
\begin{equation}
 \label{equ:pore_throat_area}
 A_n = A_{\Delta ABC} - A _{lb} - 0.5R_1^2 \alpha - 0.5R_2^2 \beta - 0.5R_3^2 \gamma 
\end{equation}
Combining Eq.(\ref{equ:capillary_pressure}), (\ref{equ:entry_force}) and (\ref{equ:pore_throat_area}), an explicit expression of $F^p$ is obtained.

We now evaluate the force coming from the interfaces. Since $\theta=0$ in Eq.(\ref{equ:tension_force_simplify1}) for perfect wetting, the total force on the interface line is
\begin{equation}
 \label{equ:tension_force_simplify2}
 T^{\gamma}(r_c)=(L_{nw}+L_{ns})\gamma^{nw}
\end{equation}
The contact lines $L_{nw}$ and $L_{ns}$ can be obtained as follows:
\begin{equation}
 \label{equ:length_nw}
 L _{nw} = r_{c} \varphi _{11} + r_{c} \varphi _{22} + r_{c} \varphi _{33}
\end{equation}
\begin{equation}
 \label{equ:length_ns}
 L _{ns} = R_{1}(\alpha - \varphi _{21} - \varphi _{31}) + R_{2}(\beta - \varphi _{32} -\varphi _{12}) + R_{3}(\gamma - \varphi _{13} - \varphi_{23})
\end{equation}
Combining equations (\ref{equ:tension_force_simplify2}), (\ref{equ:length_nw}) and (\ref{equ:length_ns}) gives an explicit expression of $T^{\gamma}(r_c)$.

\subsection{Determination of lower and upper bounds of $r_c$}
For a particular geometry, we define lower and upper bounds for $r_c$, denoted by $[r_{min},r_{max}]$. The maximum value $r_{max}$ is defined as the radius the circle inscribed between the solid particles, i.e. the solution of Apollonius's problem in Euclidean plane geometry. We solve for $r_{max}$ by using the algorithm of \cite{Chareyre2011}.

The minimum value $r_{min}$ is locally determined by the maximum distance between two neighboring particles, which is obtained by:
\begin{equation}
 \label{equ:r_min}
 r_{min} = \frac{1}{2}max\{(a-R_2-R_3),(b-R_1-R_3),(c-R_1-R_2)\}
\end{equation}
The bounds are used to initialize an iterative algorithm (dichotomy) which approximates $r_{c,e}$ the value of $r_c$ which satisfies Eq. 10. In a few cases (very flat triangles, for instance), it can happen that the solution is out of the bounds, in which case we retain $r_{min}$ to evaluate $r_{c,e}$.


\bibliographystyle{abbrv}
\bibliography{paper1arXiv.bbl}

\end{document}